\documentclass[fleqn,twoside]{article}
\usepackage{amssymb}
\usepackage{espcrc2}
\usepackage{graphicx}

\newcommand\eg{{\it e.g.}}
\newcommand\etal{{\it et al.}}
\newcommand\psihat{{\hat \psi}}
\newcommand\psibar{{\overline\psi}}
\newcommand\Sigmahat{{\hat \Sigma}}
\newcommand\sigmahat{{\hat \sigma}}
\newcommand\psibarhat{{\hat \psibar}}
\newcommand\slashnext[1]{\mathpalette{\bgroup\let\style=}
                                 {\setbox0=\hbox{$\style #1$}%
                                  \setbox2=\hbox to\wd0{\hss$\style/$\hss}%
                                  \wd2=0pt\dp2=0pt\box2\box0\egroup}}
\newcommand\Dslash{\mathop{\slashnext D}}
\newcommand\pslash{{\slashnext p}}
\newcommand\delslash{\mathop{\slashnext\partial}}

\title{Renormalization constants using quark states in Landau gauge\thanks{This research
       is supported by Department of Energy, under contract W-7405-ENG-36 and the Grand
       Challenges award at ACL at Los Alamos.}}

\author{T. Bhattacharya\address{MS B285, Group T-8, Los Alamos National
        Laboratory, Los Alamos, NM 87545, USA},
        R. Gupta\addressmark,
        W. Lee\addressmark}
       
\begin{document}

\begin{abstract}
We show that given one \(O(a)\) improvement constant, \(b_m\), all the
remaining quantities needed to define the renormalized and \(O(a)\)
improved dimension-3 quark bilinears can be obtained by studying the
matrix elements of these operators between external quark states in a
fixed gauge.
\end{abstract}

\maketitle

\section{INTRODUCTION}

One of the leading uncertainties in lattice calculations involves the
connection between the lattice and the continuum renormalized
operators.  Current estimates~\cite{us} show that one-loop
perturbation theory for $O(a)$ improved Wilson fermions underestimates
quantities like $Z_P^0/Z_S^0$ by $\sim 10\%$ at lattice scales of
$2$--$4$ GeV.  Furthermore, perturbative estimates of the $O(a)$
improvement coefficients are significantly different from their
non-perturbative estimates.

All the scale independent renormalization constants and the
improvement coefficients in the quenched theory have been determined
by imposing vector and axial ward identities on the lattice.  This
Ward identity method is computationally intensive and alternate
methods are therefore desirable, especially to determine the scale
dependent renormalization constants ($Z_T^0$ and $Z_P^0$ or $Z_S^0$).

An alternate well known method that gives all the renormalization
constants~\cite{oldrenorm} and some of the improvement
constants~\cite{martinelli,mytalk} involves calculating the matrix
elements of the quark bilinears between external quark states in a
fixed gauge. This method is far more tractable computationally and the
generalization to 4-fermion operators is straightforward.  Here, we
show how this method gives all but one ($b_m$ as discussed in
section~\ref{sec:31}) of the \(O(a)\) renormalization constants.

\vskip-0.5\baselineskip\hrule width 0pt

\section{METHOD}

We start by defining the notation.  \(O(a)\) improvement of the theory
is achieved by improving the action and the operators simultaneously.
The improved renormalized quark fields, \(\psihat\) can be
related to the lattice quark field \(\psi\) by
\begin{eqnarray}
\psihat &=& Z_\psi^{-1/2} (1 - b_\psi a {m_I}) \times{} \nonumber\\
    && \qquad [ 1 - a c'_\psi (\Dslash + {m_I})
                                             - a c_{NGI} \delslash ]
                              \psi\,,
\end{eqnarray}
where \(m_I\) is an \(O(a)\) improved quark mass.  Thus, apart from a
mass dependent renormalization constant, one needs (i) an equation of
motion correction, $ c'_\psi$, that does not affect position space
correlation functions at finite separations, and (ii) mixing with a
gauge non-invariant operator, $c_{NGI}$, that appears because the
calculation is performed in a fixed gauge.~\cite{martinelli}

We also write the renormalized propagator as
\begin{equation}
\langle \mathop{\rm T} [\psihat \psibarhat] \rangle \equiv 
   \frac{1}{i \Sigmahat_1 \pslash + \Sigmahat_2} \equiv
   -i \sigmahat_1 \pslash + \sigmahat_2\,,
\end{equation}
where \(\mathop{\rm T}\) is the time-ordering symbol, and \(p_\mu\) is
the four momentum of the quark.  Neglecting logarithmic and higher
order corrections in \(p^{-2}\), the improved renormalized propagator
at high momenta is given by
\begin{eqnarray}
\Sigmahat_1 &=& 1 + \frac{\alpha_1 {m_I}^2 + \beta_1}{p^2} 
                 \label{sigh1}\\ 
\Sigmahat_2 &=& Z_m {m_I} \left[ 1 + \frac{\alpha_2 {m_I}^2 +
                 \beta_2}{p^2} \right] + \frac{\beta'_2}{p^2}
	 \,,\label{sigh2}
\end{eqnarray}
where \(Z_m\) is the renormalization constant of $m_I$.  The important
point to note in these expressions is that chiral symmetry prevents
terms proportional to \({m_I}\) and \({m_I}^2\) in \(\Sigmahat_1\) and
\(\Sigmahat_2\) respectively.

\begin{figure}[t]
\includegraphics[width=0.9\hsize]{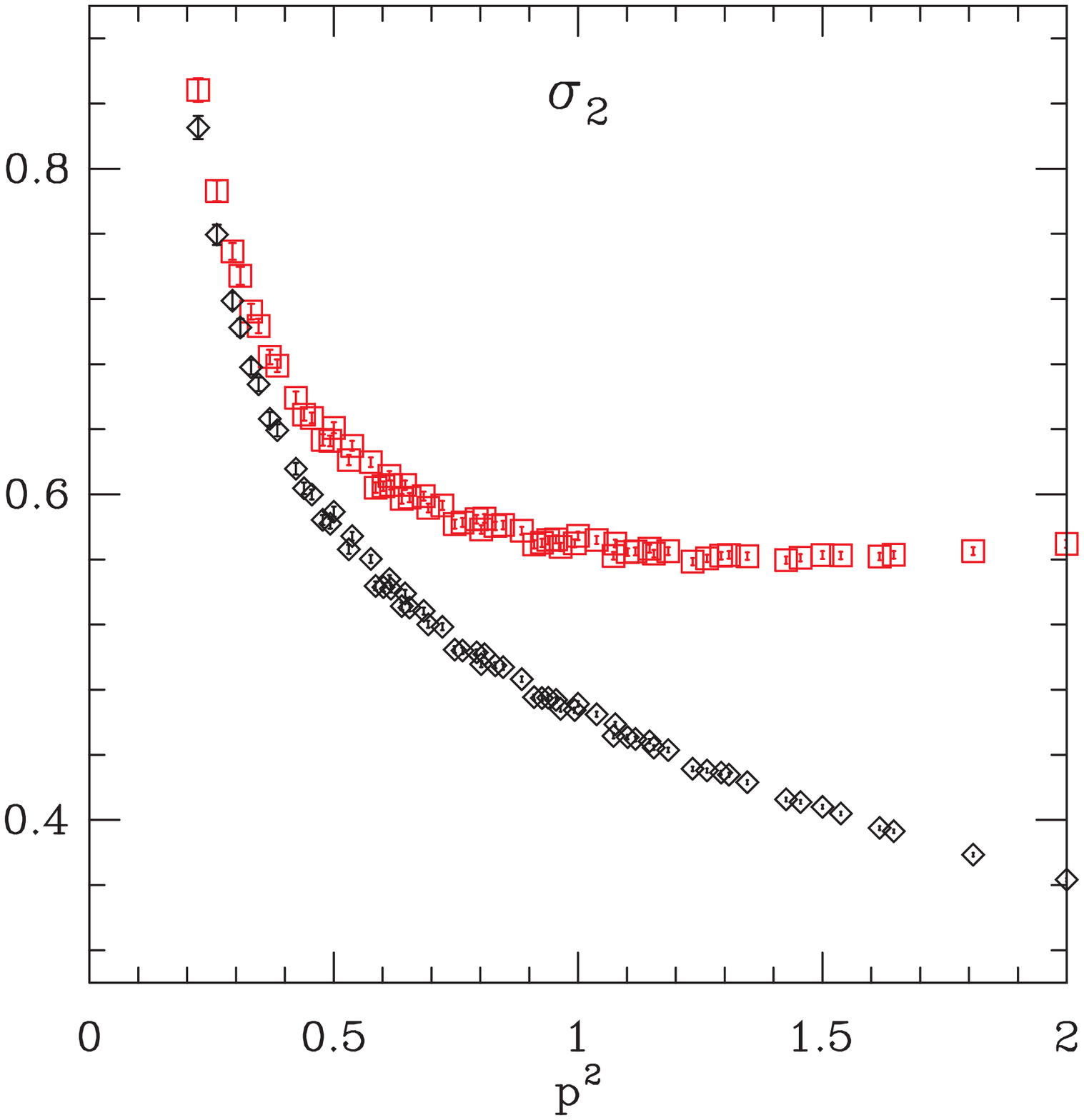}
\vspace{-2\baselineskip}
\caption{Plot of \(\sigma_2\) versus \(p^2\) before (diamonds) and
after (squares) subtraction of \(O(a^2p^2)\) artefact for \(\kappa=0.1344\).}
\label{oap2}
\end{figure}

Expanding the lattice quark fields in terms of the continuum field, 
lattice propagator is 
\begin{eqnarray}
\langle \mathop{\rm T} [ \psi \psibar ] \rangle 
     &\equiv& - i \sigma_1 \pslash + \sigma_2 \label{latsigdef}\\
     &=& Z_\psi ( 1 + 2 b_\psi a m ) (1 + 2 i a c_{NGI} \pslash)
           \times {} \nonumber\\
     &&\qquad   (- i \sigmahat_1 \pslash + \sigmahat_2) + 2 a c'_\psi
     \,.   \label{latcontsigrel}
\end{eqnarray}
From this the unknown constants \(Z_\psi\), \(Z_m\), \(b_\psi\),
\(c_{NGI}\), and \(c'_\psi\) can be extracted as follows.  We
first expand \(\sigma_1\) and \(\sigma_2\) at large \(p^2\) as
\begin{eqnarray}
p^2 \sigma_1 &=& {\sigma_1^{LO}} + \frac{\sigma_1^{NLO}}{p^2} +
                           O(p^{-4}) \label{sig1d}\\
\sigma_2 &=& {\sigma_2^{LO}} + \frac{\sigma_2^{NLO}}{p^2} + O(p^{-4})\,.
                           \label{sig2d} 
\end{eqnarray}
where the terms dropped are yet higher order in \(p^{-2}\).  From
Eqns.~\ref{sigh1}--\ref{sig2d}, we note
that these leading and next to leading coefficients,
\(\sigma^{(N)LO}_{1,2}\), of the expansion of \(\sigma_{1,2}\) in
\(p^{-2}\) have the following dependence on the quark mass:
\begin{eqnarray}
\sigma^{LO}_1  &=& \sigma^ {LO,0}_1 + \sigma^ {LO,1}_1 {m_I} \\
\sigma^{NLO}_1 &=& \sigma^{NLO,0}_1 + \sigma^{NLO,1}_1 {m_I} + {}\nonumber\\
               &&\qquad  \sigma^{NLO,2}_1 {m_I}^2 + \sigma^{NLO,3}_1 {m_I}^3 \\
\sigma^{LO}_2  &=& \sigma^ {LO,0}_2 \\
\sigma^{NLO}_2 &=& \sigma^{NLO,0}_2 + \sigma^{NLO,1}_2 {m_I} + {}\nonumber\\
               &&\qquad   \sigma^{NLO,2}_2 {m_I}^2 \,,
\end{eqnarray}
where, omitting terms of \(O(a^2)\), 
\begin{eqnarray}
\sigma^{LO,0}_1  &=& Z_\psi \label{expansion0}\\
\sigma^{LO,1}_1  &=& 2 a Z_\psi (b_\psi - c_{NGI}) \label{expansion1}\\
\sigma^{NLO,0}_1 &=& - 2 Z_\psi (\beta_1 - 2 a c_{NGI} \beta'_2) \label{expansion2}\\
\sigma^{NLO,1}_1 &=& - 2 a Z_\psi [\beta_1 b_\psi - c_{NGI} (2\beta_1
                                        - \beta_2)] \label{expansion3}\\
\sigma^{NLO,2}_1 &=& - Z_\psi (1 + \alpha_1) \label{expansion4}\\
\sigma^{NLO,3}_1 &=& - a Z_\psi [b_\psi (1 + c_{NGI}) - {}\nonumber\\
                 && \qquad 2 c_{NGI} (1 +
                           2\alpha_1 - \alpha_2)] \label{expansion5}\\
\sigma^{LO,0}_2  &=& 2 a Z_m ( Z_\psi c_{NGI} + c'_\psi ) \label{expansion6}\\
\sigma^{NLO,0}_2 &=& 2 a Z_m Z_\psi c_{NGI} \beta_1 \label{expansion7}\\
\sigma^{NLO,1}_2 &=& Z_m Z_\psi \label{expansion8}\\
\sigma^{NLO,2}_2 &=& 2 a Z_m Z_\psi [ b_\psi + c_{NGI} (1 + \alpha_1) ] \,.
\label{expansion9}
\end{eqnarray}
The five renormalization and improvement constants can now be obtained
once \(\alpha_1\) is eliminated using \(\sigma^{LO,0}_1\),
\(\sigma^{LO,1}_1\), \(\sigma^{NLO,2}_1\), \(\sigma^{LO,0}_2\),
\(\sigma^{NLO,1}_2\), and \(\sigma^{NLO,2}_2\). Once the improved
propagator is known, the remaining constants are obtained as discussed
in~\cite{mytalk}. 

\begin{figure}[t]
\includegraphics[width=0.9\hsize]{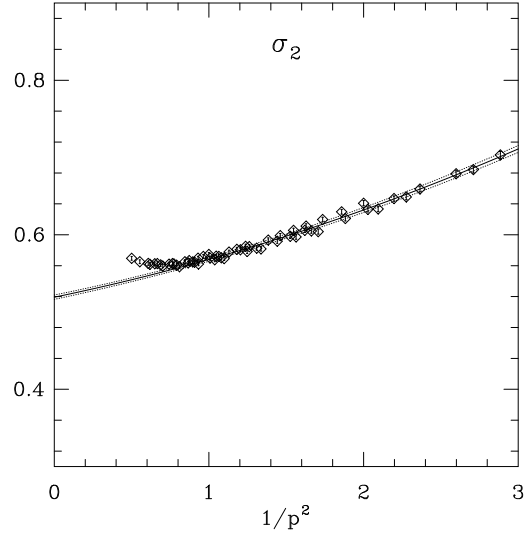}
\vspace{-2\baselineskip}
\caption{Plot of \(\sigma_2\) versus \(p^{-2}\) after
subtraction of \(O(a^2p^2)\) artefact for \(\kappa=0.1344\).} 
\label{p2fit}
\end{figure}

\section{EXAMPLE IMPLEMENTATION}

\subsection{Lattice details}
\label{sec:31}

To illustrate the feasibility of this method, we present a preliminary
analysis of 50 \(32^4\) quenched configurations at \(\beta=6.2\).  We
take \(c_{SW} = 1. 614\) from the ALPHA collaboration~\cite{Luscher}
to define the $O(a)$ improved action.  Simulations are done at seven
values of \(\kappa = 0.131\), 0.1321, 0.1333, 0.1339, 0.1344, 0.1348,
and 0.1350.  The critical value of the hopping constant, \(\kappa_c =
0.135899\), and the fourth root of the plaquette, \(u_0 = 0.88510\)
are taken from~\cite{us}.  For the \(O(a)\) improved definition of
quark mass we take\looseness-1
\begin{equation}
{m_I} = \ln \left[ 1 + \left( \frac{1}{2\kappa u_0} - \frac{1}{2\kappa_c
                                                            u_0}
                                                            \right)\right]
\end{equation}
since the value of \(b_m\) is close to its tadpole improved tree level
value~\cite{us}. We scale all lattice fermion fields by
\(\sqrt{2\kappa u_0}\). The lattice momentum components are \(p \equiv
\sin (2\pi j/32) \) and we average over momentum combinations
equivalent under the hypercubic lattice symmetry group. Our fits use
the seventy distinct momentum combinations with \(j \leq
4\).\looseness-1

\subsection{\relax\(O(a^2)\) subtraction and determination of the
constants} 

According to Eqs.~\ref{sig1d} and \ref{sig2d}, \(p^2 \sigma_1\) and
\(\sigma_2\) are supposed to go to constants at large \(p^2\),
however, the data, exemplified in Figure~\ref{oap2}, show a linear
behavior in \(a^2 p^2\) at large \(p^2\).  These \(O(a^2)\) terms are
removed by fitting to the large momentum behavior.

\begin{figure}[t]
\includegraphics[width=0.9\hsize]{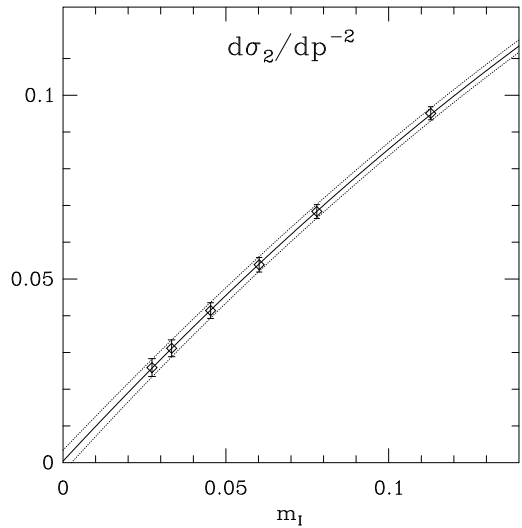}
\vspace{-2\baselineskip}
\caption{Slope of \(O(a^2)\)-corrected \(\sigma_2\) with respect to
\(p^{-2}\) plotted versus \(m_I\).}
\label{mfit}
\end{figure}

To extract the desired coefficients \(\sigma^{(N)LO}_{1,2}\) we now
fit these subtracted \(p^2 \sigma_1\) and \(\sigma_2\) as a function
of \(p^{-2}\) (see Figure~\ref{p2fit} for an example).  The data show
that one needs to keep at least terms up to \(p^{-4}\) in
Eqs.~\ref{sig1d},\ref{sig2d} to obtain a reasonable fit.  The resulting 
intercepts and slopes obtained are then fit against \({m_I}\)
(\eg, see Figure~\ref{mfit}) to obtain the various expressions defined
in Eqns.~\ref{expansion0}--\ref{expansion9}.  From these fits, we find
\begin{eqnarray}
Z_\psi  &=& 0.925(6) u_0\\
Z_m     &=& 1.03(4) u_0^{-1} \\
c_{NGI} &=& - 0.02(5) \\
b_\psi  &=& - 0.53(3) u_0^{-1} \\
c'_\psi &=& 0.27(4)
\end{eqnarray}

\vspace{-\baselineskip}
\section{DISCUSSION}

We have shown that once \(b_m\), or equivalently, \({m_I}\) is known,
all other \(O(a)\) improvement constants needed to define the
renormalized propagator can be determined.  In previous
work~\cite{martinelli}, the constant \(c_{NGI}\) was left
undetermined. Using perturbation theory Sharpe~\cite{sharpe} has shown
that the effect of this term is small.  We show that this constant can
be determined non-perturbatively and its value is indeed small.

\end{document}